\documentclass[aps,prl,twocolumn,showpacs]{revtex4}
\usepackage{amssymb,amsmath}
\usepackage{bm, graphicx, amsmath}
\usepackage{bbm}
\usepackage[section]{placeins}

\usepackage[mathscr]{eucal}
\usepackage{graphicx}
\usepackage{color}

  \newcommand{\ke}[1]{|#1\rangle}
  \newcommand{\bk}[2]{\langle #1|#2\rangle}

\begin{document}

  \title{Probabilistically Perfect Cloning of Two Pure States: A Geometric Approach}
\author{V. Yerokhin$^{1}$, A. Shehu$^{1}$, E. Feldman$^{2}$, E. Bagan$^{1,3}$ and J. A. Bergou$^{1}$}
\affiliation{$^{1}$Department of Physics and Astronomy, Hunter College of the City University of New York, 695 Park Avenue, New York, NY 10065, USA\\
$^{2}$Department of Mathematics, Graduate Center of the City University of New York, 365 Fifth Avenue, New York, New York 10016, USA \\
$^{3}$F\'{i}sica Te\`{o}rica: Informaci\'{o} i Fen\`{o}mens Qu\`antics, Universitat Aut\`{o}noma de Barcelona, 08193 Bellaterra (Barcelona), Spain
}

\begin{abstract} 
We solve the long-standing problem of making $n$ perfect clones from $m$ copies of one of two known pure states with minimum failure probability in the general case where the known states have arbitrary {\emph{a priori}} probabilities.  
The solution emerges from a geometric formulation of the problem. This formulation also reveals a deeper connection between cloning and state discrimination. The convergence of cloning to state discrimination as the number of clones goes to infinity exhibits a phenomenon analogous to a second order symmetry breaking phase transition.
\end{abstract}
\maketitle 

It is impossible to always clone quantum states perfectly \cite{Wooters, Dieks}.  This leads to advantages for quantum systems over classical ones in quantum based communications protocols of practical relevance.  A common example is provable security in quantum key distribution, with recent works showing more applications~\cite{Pomarico, Bart}.  Developments in cloning, including deterministic but approximate cloning~\cite{Buzek1,Gisin1,Buzek2,Brub,Chefles1,Fiurasek} and probabilistically perfect cloning \cite {DuanGuo,Fiurasek1,Muller} provide anchors for better understanding quantum theory as a whole, such as the relationship between the no-cloning and no-signaling theorems~\cite{Barnum}, and  fundamental limits on quantum measurements~\cite{Chiribella,Bae,Chiribella2006,Gendra}. In particular, cloning's relationship with the fundamental limits of state discrimination will be central to this Letter. For reviews citing recent developments, applications and experiments related to cloning see~\cite{review1,Fan} and to state discrimination see \cite{Bergou}.

When knowledge of the state preparation is available, perfect cloning is probabilistically possible. With the first result in probabilistic cloning, Duan and Guo \cite{DuanGuo} considered the problem of producing perfect clones of linearly independent pure states and focused on the two state case.  They found the maximum average success rate when both states are equally likely, the case of equal priors, and set this success probability as lower bound for arbitrary prior probabilities.  While other work has been done on this problem, there has until now been no general solution. In this Letter we obtain the general analytic solution and, with its help, examine the relationship of cloning and state discrimination.
 
There are a number of reasons why one might want to solve the general problem with arbitrary priors.  (i)~The solution to the equal prior problem is obtained using only symmetry arguments, with no need for optimization. (ii)~A general solution would check the robustness of the case of equal priors against variations of the prior probabilities around $1/2$.  This gives control over errors that are unavoidable for any physical realization. (iii)~One could consider a discrimination protocol consisting of optimal cloning followed by optimal Unambiguous Discrimination (UD) of the produced clones. This is a particular case of a ``prepare and measure protocol" which we will call ``discrimination by cloning." Surprisingly, this is optimal for equal priors and for any number of clones, $n$, as shown later.  This suggests that the case of equal priors is very special and can provide a deceptive view of cloning.  (iv)~In the limit of infinitely many clones, the optimal strategy prepares the clones according to the outcomes of UD of the input states. This is a particular case of a ``measure and prepare" protocol, which we will call ``cloning by discrimination." Since the UD measurement varies over the range of prior probabilities (a 3-outcome generalized measurement vs. a 2-outcome projective measurement, \cite{Bergou}), this hints at a similar situation for optimal cloning that can only be revealed by solving the general problem. 

Our solution shows that discrimination by cloning as outlined in (iii) is sub-optimal for unequal prior probabilities (unless one state is never sent). This indicates that the equal prior case is not representative of state dependent cloning.  Additionally, contrary to the suggestions in~(iv) above, our solution leads to a failure probability that is a smooth function of the priors. However,  the strategy converges to cloning by discrimination as~$n\to\infty$, implying a discontinuous second derivative and revealing a phenomenon similar to a second order symmetry breaking phase transition.

We envision a state dependent probabilistic cloner as a machine with an input port, an output port and two flags that herald the success or failure of cloning.  The input $|\psi_i^m\rangle=|\psi_i\rangle^{\otimes m}$, $i=1,2$ ($m$ identical copies of either $|\psi_1\rangle$ or $|\psi_2\rangle$) is fed through the input port for processing. In case of success, $n$ perfect clones~$|\psi_i^n\rangle=|\psi_i\rangle^{\otimes n}$ are delivered through the output port with probability $p_i$, conditioned on the input state being $|\psi^m_i\rangle$. Otherwise, the output is in a generic failure state, with a failure probability~$q_i=1-p_i$.

For cloning, optimality is usually addressed from a Bayesian viewpoint that assumes the states to be cloned are prepared with some prior probabilities $\eta_1$ and $\eta_2$, $\eta_1+\eta_2=1$. Then a natural cost function for the probabilistic cloning machine is the average failure probability, 
\begin{equation}
Q=\eta_1 q_1+\eta_2 q_2.
\label{obj fun}
\end{equation}
The aim is to find the optimal cloner that minimizes the cost function $Q$, and yields the minimum average failure probability $Q_{\rm min}$ for arbitrary priors $\eta_1$ and $\eta_2$.

In our formulation, similar to that in~\cite{DuanGuo}, the Hilbert space ${\mathscr H}^{\otimes m}$ of the original $m$ copies is supplemented by an ancillary space~${\mathscr H}^{\otimes(n-m)}\otimes {\mathscr H}_F$ that accommodates the additional $n-m$ clones and the success/failure flags. Next, we introduce a unitary transformation~$U$ via
\begin{equation}
U|\psi^m_i\rangle|0\rangle= \sqrt{p_i}|\psi^n_i\rangle|\alpha_i\rangle +\sqrt q_i |\Phi^{n}\rangle |\alpha_{0}\rangle,\quad i=1,2. \label{Ui}
\end{equation}
Here the ancillas are initialized in a reference state~$\ke 0$. The states of the flag associated with successful cloning, $\ke {\alpha_i}$, are orthogonal to the state associated with failure, $\ke{\alpha_{0}}$, and $\ke{\Phi^{n}}$ is a generic failure state in ${\mathscr H}^{\otimes n}$, the same for both inputs for optimality. 

We note that, although $|\alpha_1\rangle=|\alpha_2\rangle$ for optimal cloning, we consider a more general scenario allowing for $|\alpha_1\rangle \neq |\alpha_2\rangle$ to include ``cloning by discrimination." In this protocol we employ UD to identify the input state and then prepare clones of the identified state. For UD the success flag states must be fully distinguishable, so $\langle\alpha_1|\alpha_2\rangle=0$. Further, an even more general scenario could allow for two different failure states, $|\Phi_i^{n}\rangle$ ($i=1,2$), in Eqs.~(\ref{Ui}). This, however, would necessarily be sub-optimal since we could probabilistically determine whether we received~$\ke{\psi_1}$ or $\ke{\psi_2}$ by applying UD to the states $\ke {\Phi_i^{n}}$.  In case of success, we could prepare $n$ copies of the state, increasing the overall success rate of the cloning strategy.

To proceed, we now take the inner product of equation~$i$ with itself in \eqref{Ui}, yielding that the probabilities are normalized, $p_i+q_i=1$. Taking the inner product of equation $i=1$ with $i=2$ in \eqref{Ui} yields the main constraint, 
\begin{equation}
s^m=\sqrt{p_1 p_2}\, s^n \alpha+\sqrt{q_1 q_2},
\label{unit cond}
\end{equation}
which is a consequence of the unitarity of $U$. Here we introduced the notation $s \equiv \bk {\psi_1}{\psi_2}$ and  $\alpha \equiv \langle\alpha_1|\alpha_2\rangle$. The overlaps $s$ and $\alpha$ can be chosen to be real without any loss of generality, so $0 \le s, \alpha \le 1$.  Clearly, $\alpha=1$ for optimal cloning, while~$\alpha=0$ for cloning by discrimination. If Eq.~(\ref{unit cond}) is satisfied, it is not hard to prove that~$U$ has a unitary extension on the whole space.  

\begin{figure}[h,t]
\centering
\includegraphics[width=14em]{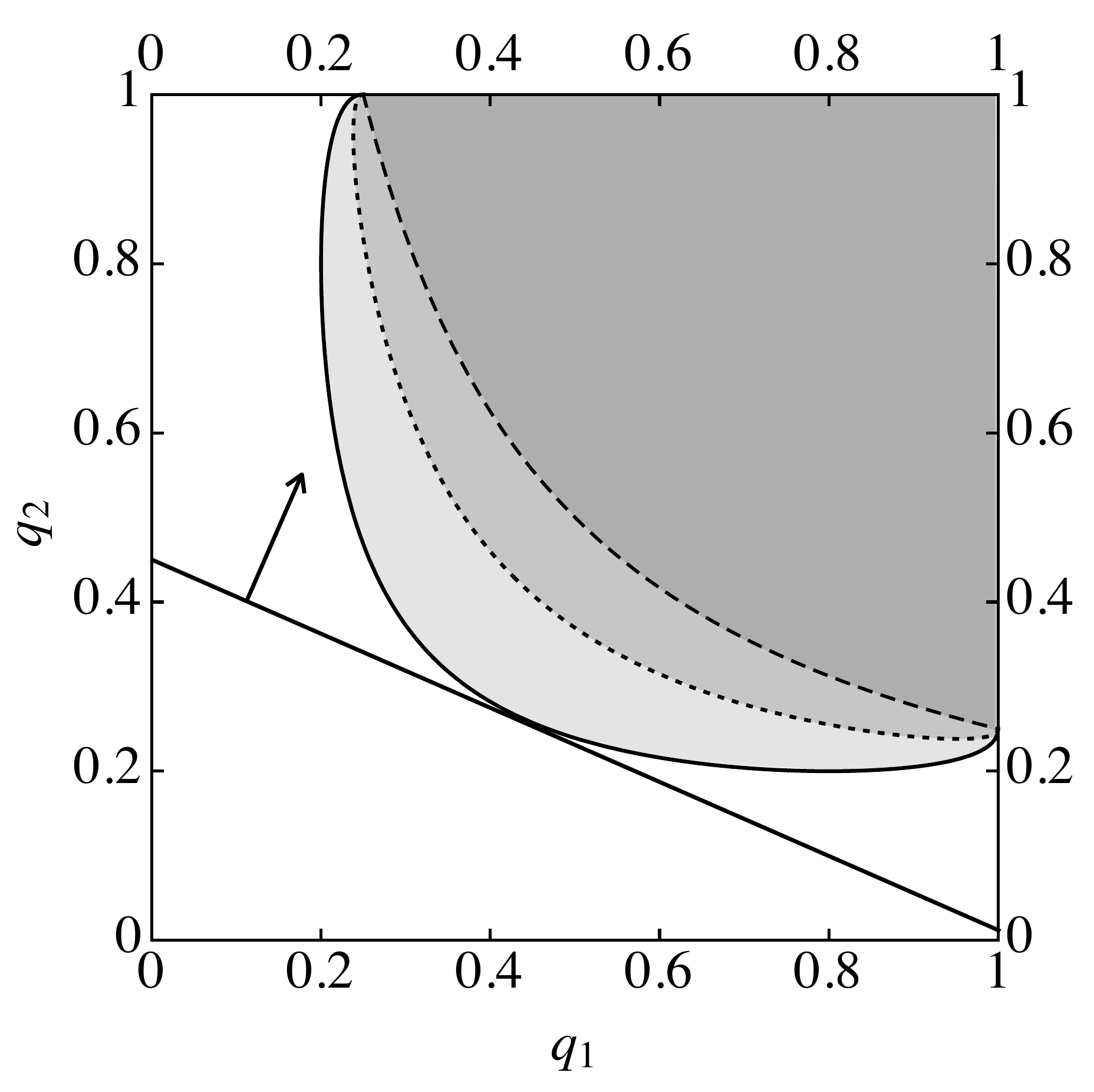}
\caption{Unitarity curves, $q_{2}$ vs. $q_{1}$, from Eq.~(\ref{unit cond}) and the associated sets~$S_\alpha$ from Eq.~(\ref{S_alpha}) for $\alpha=1$ (solid/light~gray), $\alpha=0.5$ (dotted/medium gray), and $\alpha=0$ (dashed/dark gray). The figure also shows the optimal straight segment \mbox{$Q=\eta_1 q_1+\eta_2 q_2$} and its normal vector~$(\eta_1,\eta_2)$. For all plots~$s = 0.5$, $m = 1$, and $n =2$.}
\label{fig:1}
\end{figure}

Before developing the analytical theory of optimizing (minimizing) $Q$, we present a complete geometric picture, similar in spirit to that in~\cite{Bergou1}, that visually solves the optimization problem and serves as guide for the subsequent calculations. 
For the geometrization the following features of \eqref{unit cond} turn out to be important. (a)~For fixed $s$, $n$ and $m$ \eqref{unit cond} defines a class of smooth curves on the unit square $0\le q_i\le 1$ (e.g., solid, dashed or dotted curves in Fig.~\ref{fig:1}). (b)~All these curves meet at their endpoints, $(1,s^{2m})$ and $(s^{2m},1)$. (c)~At the endpoints  the curves become tangent to the vertical and horizontal lines~$q_1=1$ and~$q_2=1$, respectively, provided~$\alpha\not=0$. 
(d)~For~$\alpha=0$ the curve defined by Eq.~(\ref{unit cond}) is the hyperbola $q_1 q_2=s^{2m}$ (dashed line in Fig.~\ref{fig:1}). (e)~The curve corresponding to a particular value of $\alpha$ and the segments joining its end points with the vertex~$(1,1)$ form the boundary of the set $S_{\alpha}$ (any of the gray regions in Fig.~\ref{fig:1}), where
\begin{equation}
S_\alpha=\{ (q_1,q_2): \sqrt{p_1 p_2}\,s^n\alpha+\sqrt{q_1 q_2}-s^m\ge 0\}.
\label{S_alpha}
\end{equation}
The sets satisfy $S_{\alpha} \subset S_{\alpha'}$ if $\alpha<\alpha'$.
(f)~The sets~$S_\alpha$ are convex if $\alpha \ge 0$. In particular $S_1$ is convex.

These considerations lead to the emergence of a geometrical picture of the optimization problem which we display in Fig.~\ref{fig:1}.

Eq.~(\ref{obj fun}) defines a straight segment on the unit square $0\le q_i\le 1$ with a normal vector in the first quadrant parallel to $(\eta_1,\eta_2)$. For fixed priors, the average failure probability~$Q$ is proportional to the distance from this segment to the origin~$(0,0)$. 
Since $S_1$ is convex and the stretch of its boundary given by Eq.~(\ref{unit cond}) with $\alpha=1$ is smooth, a unique point $(q_1,q_2)$ of tangency with the segment~(\ref{obj fun}) exists for any value of the priors and finite $n$.
It gives $Q_{\rm min}$ and defines the optimal cloning strategy. 

We note that the inclusion hierarchy of the sets $S_\alpha$ provides a simple geometrical proof that $\alpha=1$, i.e., $|\alpha_1\rangle=|\alpha_2\rangle$, is the optimal choice for cloning. For ``cloning by discrimination," on the other hand, $\langle\alpha_1|\alpha_2\rangle=\alpha=0$. From points~(d) and~(e) above, it follows that for any finite $n$ this protocol is strictly suboptimal, i.e., $Q_{\rm min}<Q_{\rm UD}$, noticing that the failure rate of cloning by discrimination is that of UD, $Q_{\rm UD}$. This is intuitively obvious since in cloning one is asking for less than in UD; the identity of the input states is not revealed for any finite~$n$. However, optimal cloning and UD become one and the same in the limit $n \to \infty$, when $s^n \to 0$ and the curve~(\ref{unit cond}) collapses to the hyperbola $q_1 q_2=s^{2m}$, as it does for $\alpha=0$. We return to this point below.

\begin{figure}[hh]
\centering
$%
\begin{array}{c}
\includegraphics[width=26.8em]{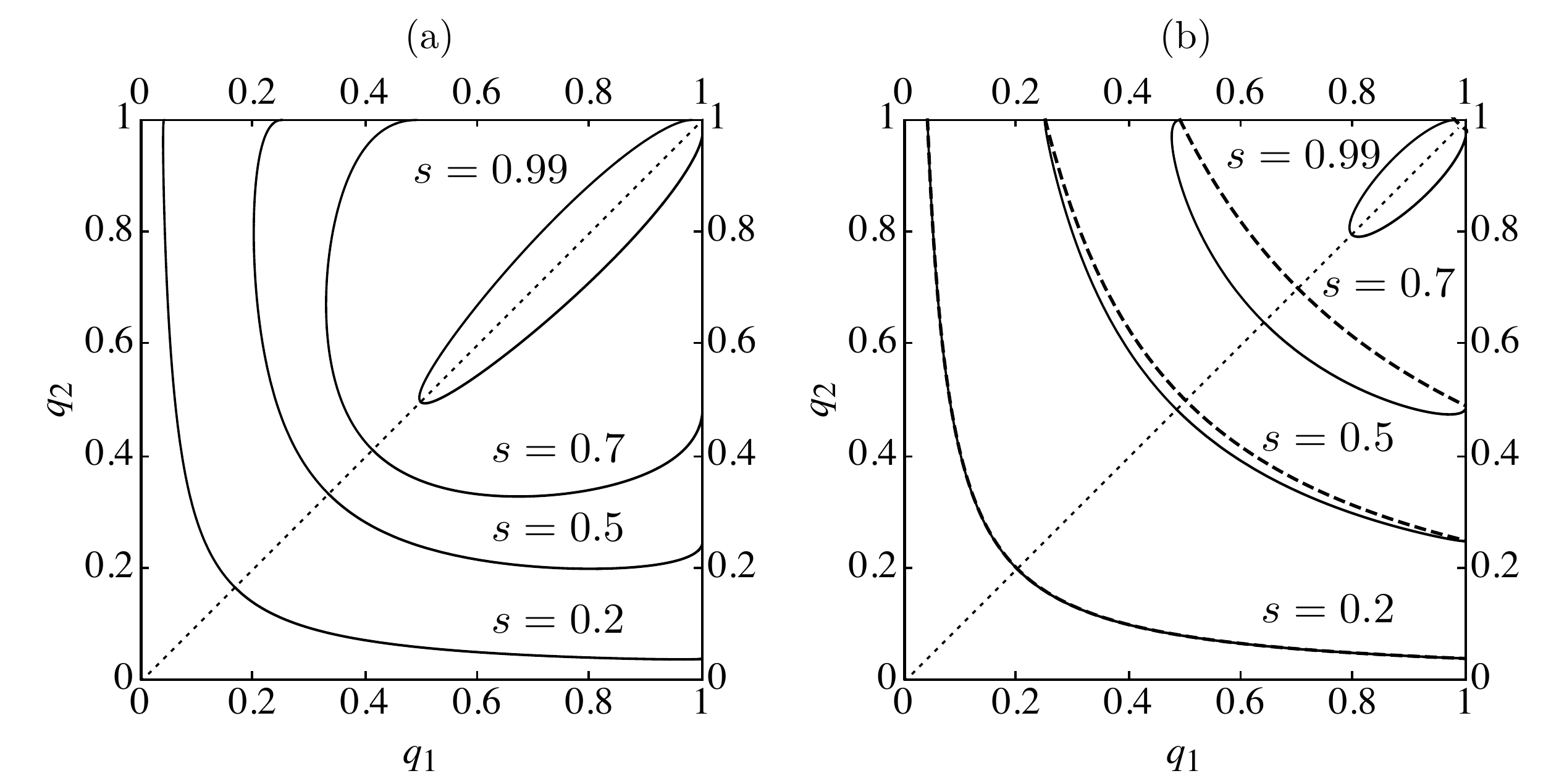}\\
\end{array}%
$%
\caption{Unitarity curves, $q_{2}$ vs. $q_{1}$, from Eq.~(\ref{unit cond}) for different values of~$s$ and for (a) $m=1$, $n=2$ and (b) $m=1$, $n=5$. The curves are symmetric under mirror reflection along the dotted line $q_1=q_2$, i.e., under the transformation~$q_1\leftrightarrow q_2$. The dashed lines in~(b) are the hyperbolae~$q_1 q_2=s^{2m}$.}
\label{fig:2}
\end{figure}

A more quantitative analysis requires finding a convenient parametrization of the constraint~(\ref{unit cond}). To this end, we set $\alpha=1$ and write $\sqrt{q_i} = \sin \theta_i$ for $0\leq \theta_i \leq \pi/2$. By further introducing the variables $x =\cos(\theta_1+\theta_2)$ and $y = \cos (\theta_1 - \theta_2)$ we manage to linearize the curve~(\ref{unit cond}),
which now is the straight segment $2s^m=(1+s^n)y-(1-s^n)x$ with $|x| \le y\le 1$.  
Its guiding vector is readily seen to be $(1+s^n,1-s^n)$, so the segment's parametric equation can be written as 
\begin{equation}
x=\frac{1-(1+s^n)t}{s^{n-m}},\qquad y=\frac{1-(1-s^n)t}{s^{n-m}},
\label{x & y}
\end{equation}
where we have rescaled the guiding vector so that Eqs.~(\ref{t's}) and~(\ref{qi'}) below are simplest. 
Because of the symmetry of this procedure, the parameters $x$ and $y$ are invariant under $q_1\leftrightarrow q_2$ (equivalently, under $\theta_1\leftrightarrow \theta_2$). Thus, the two mirror halves of the curve~(\ref{unit cond}) under this transformation are mapped  onto the same straight line~(\ref{x & y}). By expressing $q_i$ as a function of~$t$ only half of the original curve is recovered. The other half is trivially obtained by applying~$q_1\leftrightarrow q_2$.
After putting the various pieces together 
one  can easily get rid of the trigonometric functions and express Eq.~(\ref{unit cond}) in parametric form as 
\begin{equation}
q_i=\frac{1-xy-(-1)^i\sqrt{1-x^2}\sqrt{1-y^2}}{2} ,\quad i=1,2.
\label{par sqrt}
\end{equation}
Fig.~\ref{fig:2} shows examples of the unitary curve~(\ref{unit cond}). 

For $n > 2$ the curves closely approximate $q_1 q_2=s^{2m}$ (dashed lines) for small and moderate values of $s$, while for $s$ close to $1$ the hyperbolas remain closer to the vertex~$(1,1)$, but retain the same end points. In the limit~$n\to\infty$ all these curves become hyperbolic.

We now return to finding the minimum, $Q_{\rm min}$, of the average failure probability $Q$. Despite its apparent simplicity, this involves solving a high-order equation without a simple form. Instead, we will derive a parametric equation for $Q_{\rm min}(\eta_1)$. Along with the complete description of the unitary curve~(\ref{unit cond}), this provides a complete solution of the problem in parametric form.

With no loss of generality we may assume $\eta_1\le\eta_2$ or, equivalently,  $0\le\eta_1\le 1/2$. Then the slope of the straight line~(\ref{obj fun}), $-\eta_{1}/ \eta_{2}$, satisfies $-1\leq -\eta_{1}/\eta_{2} \leq0$. Hence, it can only become tangent to the lower half of the unitarity curve~(\ref{unit cond}) (see Fig.~\ref{fig:2}). 
Increasing $q_{1}$, the slope of this lower half increases monotonically from $-1$ at $q_1=q_2$, to $0$ before we reach the line~$q_1=1$ (assuming $n$ is finite). This follows from the properties (a)--(f) above and can be checked using  Eq.~(\ref{par sqrt}). The values of $t$ at which the slope is $-1$ and $0$ are, respectively,
\begin{equation}
t_{-1}=\frac{1-s^{n-m}}{1-s^n},\quad
t_0=\frac{1-s^{2(n-m)}}{1-s^{2n}}.
\label{t's}
\end{equation}
For any point $(q_1(t),q_2(t))$ with $t\in[t_{-1},t_0]$ there is a line $Q=\eta_1 q_1+\eta_2 q_2$ that is tangent to it, starting with $\eta_1=\eta_2=1/2$ for $t=t_{-1}$ up to $\eta_1=0$, $\eta_2=1$ for $t=t_0$. 

This observation enables us to derive the desired parametric expression for the optimality curve~$Q_{\rm min}(\eta_1)$ as follows. For a given $t$ in the range above, a necessary condition for tangency is \mbox{$\eta_1 q'_1+\eta_2 q'_2=0$}, where $q'_i=d q_i/d t$. In this equation we can solve for $\eta_1$ (or $\eta_2$) using that~$\eta_1+\eta_2=1$. By substituting $q_1$ and~$q_2$ in Eq.~(\ref{obj fun}) with~(\ref{par sqrt}) we enforce contact with the unitarity curve and obtain the expression of $Q_{\rm min}$. The final result can be cast as:
\begin{equation}
\eta_1=\frac{q'_2}{q'_2-q'_1},\;\; Q_{\rm min}=\frac{q'_2 q_1-q'_1 q_2}{q'_2-q'_1},\;\; t_{-1}\le t\le t_0,
\label{main}
\end{equation}
where $t_{-1}$, $t_0$ and $q_i$ are given in Eqs.~(\ref{t's}) and~(\ref{par sqrt}). The expressions for the derivatives $q'_i$ are
\begin{equation}
q'_i=\frac{\sqrt{q_i(1-q_i)}}{s^{n-m}}\left\{\frac{1+s^n}{\sqrt{1-x^2}}-(-1)^i\frac{1-s^n}{\sqrt{1-y^2}}\right\}.
\label{qi'}
\end{equation}

\begin{figure}[ht]
\centering
$%
\begin{array}{c}
\includegraphics[width=26.7em]{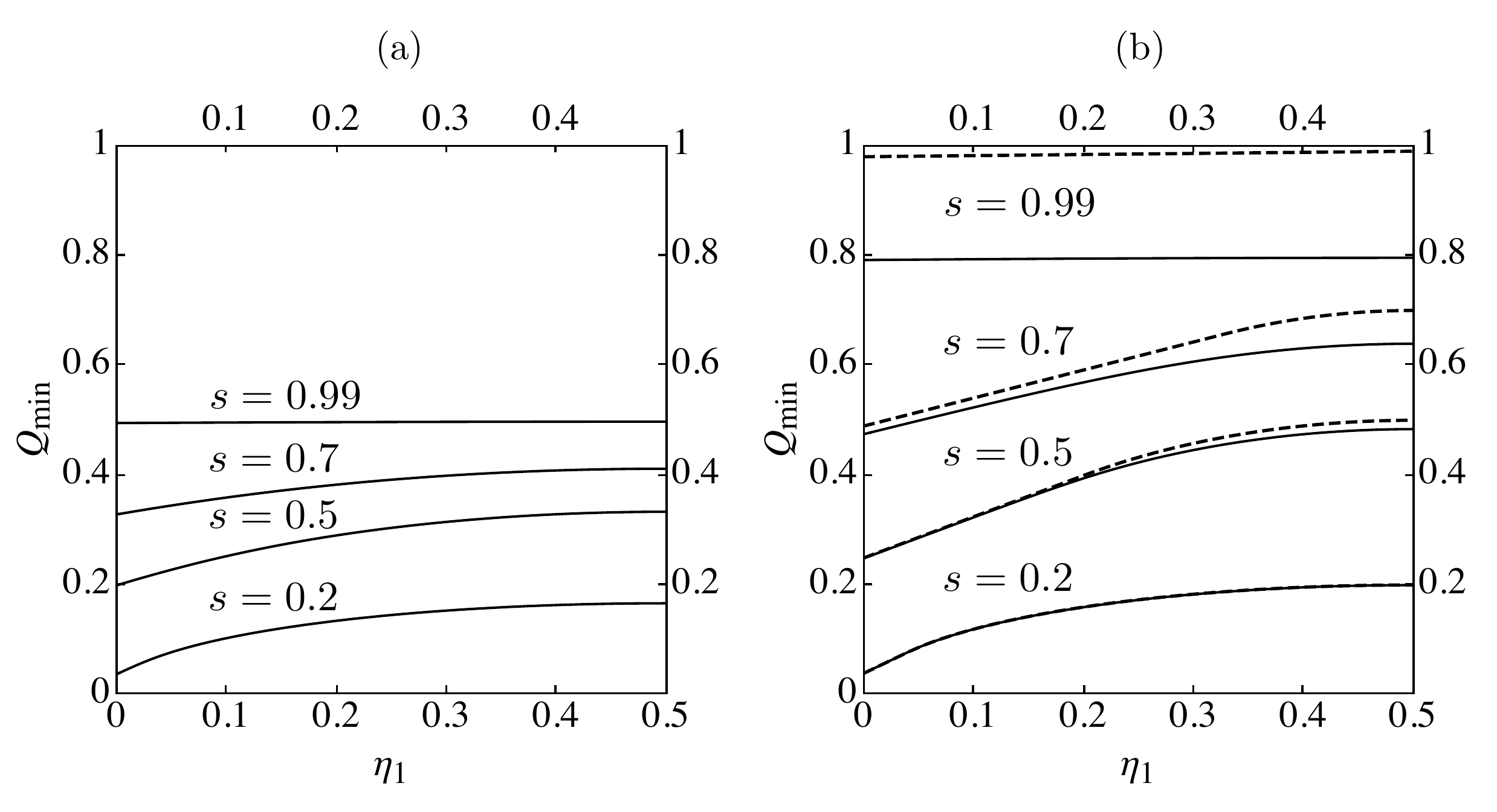}\\
\end{array}%
$%
\caption{Minimum cloning failure probability $Q_{\rm min}$ vs. $\eta_1$ (solid lines) and~UD failure probability $Q_{\rm UD}$ vs. $\eta_1$ (dashed lines) for the same values of $m$, $n$ and $s$ used in the previous figure.}
\label{fig:3}
\end{figure}

Fig.~\ref{fig:3} shows plots of the curves $Q_{\rm min}(\eta_1)$ for $m=1$ input copies and  (a) $n=2$ or (b) $n=5$ clones, as in the previous figure. We see that $Q_{\rm min}$ is an increasing function of $\eta_1$ in the given range $[0,1/2]$. The values of~$Q_{\rm min}$ at the end points of this range follow by substituting $t_{0}$ and $t_{-1}$, Eq.~(\ref{t's}), into Eq.~(\ref{par sqrt}). They are given by
\begin{equation}
Q_{0}=q_2(t_0)=\frac{s^{2m}-s^{2n}}{1-s^{2n}},\quad
Q_{-1}=\frac{s^m-s^n}{1-s^n},
\label{Q's}
\end{equation}
where $Q_{\rm min}=Q_{-1}$ holds for equal priors and $Q_{\rm min}=Q_0$ for $\eta_1\to 0$ (i.e., $\eta_2\to 1$).
The dashed lines in Fig.~\ref{fig:3}~(b) depict the well known optimal UD solution~\cite{Bergou}:
{
\begin{equation}
Q_{\rm UD}=\left\{
\begin{array}{ll}
2\sqrt{\eta_1\eta_2}\, s^m,&\displaystyle \frac{s^{2m}}{1+s^{2m}}\le\eta_1\le \frac{1}{2};\\[.5em]
\eta_1+s^{2m} \eta_2, \quad &\displaystyle 0\le \eta_1\le \frac{s^{2m}}{1+s^{2m}}.
\end{array}
\right.
\label{UD}
\end{equation}
}

It is apparent from these plots that the optimal cloning protocol performs strictly better than cloning by discrimination, as was geometricaly proved in Figs. \ref{fig:1} and \ref{fig:2}. However, the difference in performance decreases with increasing number of clones. In Fig.~\ref{fig:3}~(b), for only $n=5$, a difference is hardly noticeable for $s\le 0.5$. For $s > 0.5$ the convergence is slower but in the limit $n\to\infty$ there is perfect agreement for any $s<1$.

The convergence of the optimal cloning failure probability, $Q_{\rm min}$, to that of cloning by discrimination, $Q_{\rm UD}$ in Eq.~(\ref{UD})
follows from our geometrical approach. Recall that in the limit $n\to\infty$ (or equivalently $\alpha\to0$)  the right hand side of Eq.~(\ref{unit cond}) describes hyperbolas that we can write as $q_2=s^{2m}/q_1$ (dashed lines in Figs.~\ref{fig:1} and~\ref{fig:2}). Their  slopes are in the range $[-1,-s^{2n}]$.  A unique point of tangency with the line~(\ref{obj fun}) can only exists if the slope of this line, $-\eta_1/\eta_2$, is within this same range. This gives the $\eta_1$ interval in the first line of~(\ref{UD}), and one can easily obtain the corresponding expression for~$Q_{\rm UD}$.
If the slope of the line~(\ref{obj fun}) is outside the range, tangency is not possible, and the optimal line merely touches the end points of the hyperbolas, so the expression of~$Q_{\rm UD}$ becomes the second line of Eq.~(\ref{UD}). In geometrical terms, the straight line~(\ref{obj fun}) pivots on the end points as we vary~$\eta_1$. Furthermore, note that for the second line in~(\ref{UD}) we have $p_1=1-q_1=0$, which leads to a $2$-outcome projective measurement, as only one success flag state ($|\alpha_2\rangle$) is needed in Eqs.~(\ref{Ui}). 

Interestingly, in this limit a phenomenon analogous to a second order phase transition takes place. Our geometrical approach shows that the average failure probability $Q_{\rm min}(\eta_1)$ is an infinitely differentiable function of~$\eta_1$ for finite $n$. However, as $n$ goes to infinity (or $\alpha\to0$) the limiting function $Q_{\rm UD}(\eta_1)$ has a discontinuous second derivative. Moreover, the symmetry $q_1\leftrightarrow q_2$ breaks  in the ``phase" corresponding to the second line in Eq.~(\ref{UD}). A~similar phenomenon arises in UD of more than two pure states~\cite{Bergou1}.

It has been argued above that cloning by discrimination is strictly suboptimal (unless $n\to \infty$).
One could likewise wonder if discrimination by cloning can be optimal. On heuristic grounds, one should not expect this to be so, as cloning involves a measurement and some information can be drawn from the observed outcome. However, the equal-prior and the $\eta_1\to 0$ cases provide remarkable exceptions. For both we may write the total failure rate as $Q_{\rm C} + (1-Q_{\rm C})Q_{\rm UD}$, where C stands for cloning. For $\eta_1 = \eta_2= 1/2$, Eq.~(\ref{Q's}) implies $Q_{\rm C}=Q_{-1}$, in which case the produced $n$-clone states are equally likely. The UD of these states fails with probability $s^{n}$, as follows from Eq.~(\ref{UD}) applied to $n$ copies. The total failure rate is then $s^m$, which is the optimal UD failure rate of the original input states, Eq.~(\ref{UD}). If $\eta_1\to 0$ then only $|\psi^n_2\rangle$ is produced with non-vanishing probability and  $Q_{\rm C}=Q_{0}$. Failure in the second step (UD) is given by the bottom line in Eq.~(\ref{UD}) applied to $n$ copies. The total failure rate is $s^{2m}$, also achieving optimality. 

Using our main result in Eqs.~(\ref{main}), and~(\ref{par sqrt}) one can check that these are the only cases where discrimination by cloning is optimal. These are also the only cases where no information gain can be drawn from the cloning measurement.  This hints at how special these cases are and justifies the need of the derived solution for arbitrary priors to have a full account of two-state cloning. 

In summary, we have provided the general solution to the long-standing probabilistically perfect cloning problem for two states. The unequal prior case ($\eta_1\not=\eta_2$) uncovers remarkable phenomena that the very special equal priors case was unable to reveal.  In particular, 
the convergence of cloning to unambiguous discrimination as the number of clones becomes very large involves a discontinuity in the second derivative of the failure rate $Q_{\rm min}(\eta_1)$, corresponding to a second order symmetry breaking phase transition.
Our geometrical approach proved very helpful for both visualizing what the solution looks like qualitatively and for deriving the analytical solution. The same approach can also be applied to cloning of three or more states, deterministically approximate cloning, and other optimization problems that involve highly non-linear constraints.

\begin{acknowledgments}
\emph{Acknowledgments}. This publication was made possible through the support of a Grant from the John Templeton Foundation. The opinions expressed in this publication are those of the authors and do not necessarily reflect the views of the John Templeton Foundation. Partial financial support by a Grant from PSC-CUNY is also gratefully acknowledged. The research of EB was additionally supported by 
the Spanish MICINN, through contract FIS2013-40627-P, the Generalitat de
Catalunya CIRIT, contract  2014SGR-966, and ERDF: European Regional Development Fund.
\end{acknowledgments}


\begin{thebibliography}{99}


\bibitem{Wooters} W. K. Wooters and W. H. Zurek, Nature {\bf 299}, 802 (1982).
\bibitem{Dieks} D. Dieks, Phys. Lett. A {\bf 92}, 271 (1982).
\bibitem{Pomarico} E. Pomarico, B. Sanguinetti, P. Sekatski, H. Zbinden, and N. Gisin, Optics and Spectroscopy \textbf{111}, 510 (2011).
\bibitem{Bart} K. Bartkiewicz, A. {\v C}ernoch, K. Lemr, J. Soubusta, and M. Stobi{\' n}ska, \pra {\bf 89}, 062322 (2014).
\bibitem{Buzek1} V. Bu{\v z}ek and M. Hillery, \pra {\bf 54}, 1844 (1996).
\bibitem{Gisin1}N. Gisin and S. Massar, \prl \textbf{79}, 2153 (1997).
\bibitem{Buzek2} V. Bu{\v z}ek and M. Hillery, \prl \textbf{81}, 5003 (1998).
\bibitem{Brub} D. Bru{\ss}, D. P. DiVincenzo, A. Ekert, C. A. Fuchs, C. Macchiavello,
and J. A. Smolin, \pra {\bf 57}, 2368 (1998).
\bibitem{Chefles1} A. Chefles and S. M. Barnett, \pra \textbf{60}, 136 (1999).
\bibitem{Fiurasek} J. Fiur{\' a}{\v s}ek, S. Iblisdir, S. Massar, and N. J. Cerf, \pra \textbf{65}, 040302(R) (2002).
\bibitem{DuanGuo} L. M. Duan and G. C. Guo, \prl \textbf{80}, 4999 (1998).
\bibitem{Fiurasek1} J. Fiur{\' a}{\v s}ek, \pra \textbf{70}, 032308 (2004).
\bibitem{Muller} C. R. M{\" u}ller, C. Wittmann, P. Marek, R. Filip, C. Marquardt, G. Leuchs, and U. L. Andersen, \pra \textbf{86}, 010305(R) (2012).
\bibitem{Barnum} H. Barnum, J. Barrett, M. Leifer, and A. Wilce, \prl {\bf 99}, 240501 (2007).
\bibitem{Bae} J. Bae and A. Ac\'{\i}n, \prl \textbf{97}, 030402 (2006).
\bibitem{Chiribella} G. Chiribella, Y. Yang, and A. Yao, Nature Comm. \textbf{4}, 2915 (2013).
\bibitem{Gendra}  B. Gendra, J. Calsamiglia, R. Mu{\~ n}oz-Tapia, E. Bagan, and G. Chiribella, \prl {\bf 113}, 260402 (2014).
\bibitem{Chiribella2006} G. Chiribella and G. M. D'Ariano, \prl \textbf{97}, 250503 (2006).

\bibitem{review1} V. Scarani, S. Iblisdir, N. Gisin, and A. Ac\'{\i}n, \rmp \textbf{77}, 1225 (2005).

\bibitem{Fan} H. Fan, Y. N. Wang, L. Jing, J. D. Yue, H. D. Shi, Y. L. Zhang, and L. Z. Mu, Phys. Rep. {\bf 544}, 241 (2014).

\bibitem{Bergou} J. A. Bergou, J. Mod. Opt. {\bf 57}, 160 (2010).

\bibitem{Bergou1} J. A. Bergou, U. Futschik, and E. Feldman, \prl {\bf 108}, 250502 (2012).


\end{thebibliography}
\end{document}